# Polynomial Estimators for High Frequency Moments

Sumit Ganguly

Indian Institute of Technology, Kanpur

## Abstract

We present an algorithm for computing $F_p$, the $p$th moment of an $n$-dimensional frequency vector of a data stream, for $2 < p < \log(n)$, to within $1 \pm \epsilon$ factors, $\epsilon \in [n^{-1/p}, 1]$ with high constant probability. Let $m$ be the number of stream records and $M$ be the largest magnitude of a stream update. The algorithm uses space in bits

$$O(p^2 \epsilon^{-2} n^{1-2/p} E(p,n) \log(n) \log(nmM)/\min(\log(n), \epsilon^{4/p-2}))$$

where, $E(p,n) = (1 - 2/p)^{-1}(1 - n^{-4(1-2/p)})$. Here $E(p,n)$ is $O(1)$ for $p = 2 + \Omega(1)$ and $O(\log n)$ for $p = 2 + O(1/\log(n))$. This improves upon the space required by current algorithms [10, 5, 2, 6] by a factor of at least $\Omega(\epsilon^{-4/p} \min(\log(n), \epsilon^{4/p-2}))$. The update time is $O(\log(n))$. We use a new technique for designing estimators for functions of the form $\psi(\mathbb{E}[X])$, where, $X$ is a random variable and $\psi$ is a smooth function, based on a low-degree Taylor polynomial expansion of $\psi(\mathbb{E}[X])$ around an estimate of $\mathbb{E}[X]$.

## 1 Introduction

Massive and continuous data is generated by varied sources, such as data network switches, satellite imagery, sensor networks, web-click and transaction data, and require efficient, on-the fly analysis for early warning of critical or significant behavior. The data stream model is one of well-known computational models for massive data analysis. In this model an algorithm is typically given a relatively small amount of memory to summarize a large and (often rapidly arriving) dataset and is usually allowed a single pass (or at most a few passes) over the data. The algorithm must answer queries on the dataset for which it may only use the data summary. The model has been used in earlier work by Morris [14], Munro and Patterson [15] and Flajolet and Martin [9]. The streaming model became popular due to the seminal work of Alon, Matias and Szegedy [1] and a simultaneous growth in applications for monitoring massive data from various quarters, including networking and sensor networks. A survey is given by Muthukrishnan [32].

A data stream is viewed as a sequence of records of the form $(i, v)$, where, $i \in \{1, 2, \ldots, n\} = [n]$ and $v \in [-M, \ldots, M]$ and integral. The record $(i, v)$ changes the *frequency* $f_i$ as $f_i \leftarrow f_i + v$. The vector $f = [f_1, f_2, \ldots, f_n]^T$ is called the frequency vector of the stream. Let $m$ be the number of records appearing in the stream. Given a function of the frequency vector of the stream, the problem is to compute the function using space that is sub-linear in $n$ and $m$. The space used is one of the measures used for comparing randomized algorithm streaming algorithm. The time taken to process each stream update $(i, v)$ is another measure of a streaming algorithm and is called *update* time. Finally, the time taken to report the query is the *reporting* time and is also a measure that has been used [12].

| $F_p$ Algorithm | Space $O$(words) | Update time $O(\cdot)$ |
|---|---|---|
| AMS [1] | $n^{1-1/p}\epsilon^{-2}$ | $n^{1-1/p}\epsilon^{-2}$ |
| IW [10] | $n^{1-2/p}\epsilon^{-12}\log^{O(1)} n$ | $(\log^{O(1)} n)(\log(mM))$ |
| Hss$^a$ [5] | $n^{1-2/p}\epsilon^{-2-4/p}(\log n)(\log(mM))p^2$ | $(\log n)(\log(mM))$ |
| MW [13] | $n^{1-2/p}(\epsilon^{-1}\log(n))^{O(1)}$ | $n^{1-2/p}\text{poly}(\epsilon^{-1}\log n)$ |
| AKO$^b$ [2] | $n^{1-2/p}\epsilon^{-2-6/p}(\log n)p^2 E(p,n)$ | $\log n$ |
| BO [6] | $n^{1-2/p}\epsilon^{-2-4/p}\log(n)p^2 g(p,n)$ | $\log n$ |
| this paper | $n^{1-2/p}\epsilon^{-2}\log(n)p^2 E(p,n)/\min(\log(n),\epsilon^{4/p-2})$ | $\log n$ |

Word size is $O(\log(nmM))$ bits. $E(p,n) = (1 - 2/p)^{-1}(1 - n^{-4(1-2/p)})$. $g(p,n) = \min_{c \text{ constant}} g_c(n)$, where, $g_1(n) = \log n$, $g_c(n) = \log(g_{c-1}(n))/(1 - 2/p)$.

$^a$ Hss uses a slightly modified level mapping function, see Appendix A.
$^b$ The dependence on $\epsilon$ is improved to $\epsilon^{-2-4/p}$ [3].

Figure 1: Space requirement of algorithms for $\epsilon$-approximation of $F_p$, $p > 2$.

The $p$th moment of the frequency vector $f$ is defined as $F_p = \sum_{i\in[n]}|f_i|^p$. In this work, we consider the following problem. Given $p > 2$ and $\epsilon \in (0,1]$, design an algorithm that outputs $\hat{F}_p$ such that $\Pr\big[|\hat{F}_p - F_p| \le \epsilon F_p\big] \ge 0.6$. We will only consider single-pass streaming algorithms. Alon, Matias and Szegedy in [1] first defined this problem and presented a sublinear space algorithm that used $O(n^{1-1/p}\epsilon^{-2}\log(nmM))$ bits. The same work also showed that deterministic approximation algorithms require $\Omega(n)$ space. On the space lower bound front, Alon et.al. in [1] first showed an $\Omega(n^{1-5/p})$ bound. This was significantly improved in the elegant works of Bar-Yossef et.al. and Chakrabarti, Khot and Sun in [4, 8] to $\Omega(n^{1-2/p}\epsilon^{-2/p})$ bits. Woodruff in [17] showed a bound of $\Omega(\epsilon^{-2})$ bits for all $F_p$. Recently, Jayram and Woodruff in [11] present an improved bound of $\Omega(n^{1-2/p}\log n)$ bits. The current lower bound for space is $\Omega(\epsilon^{-2} + n^{1-2/p}\epsilon^{-2/p} + n^{1-2/p}\log n)$ bits.

Indyk and Woodruff in [10] presented the first algorithm with space $\tilde{O}(n^{1-2/p})$. Table 1 lists the published algorithms for estimating $F_p$, $p > 2$, in chronological order, along with their space and update-time requirements, where, update time is the time required to process each arriving stream record. The hierarchical sampling technique introduced by Indyk and Woodruff in [10] is essentially used by all later algorithms. The Hss algorithm [5] improved the space of the IW algorithm to $O(n^{1-2/p}\epsilon^{-2-4/p}\log(n)\log(mM))$ words. Monemizadeh and Woodruff in [13] estimate $F_p$ via an $F_2$-sampler using space $n^{1-2/p}\text{poly}(\epsilon^{-1}\log n)$; they also present multi-pass algorithms. An elegant recursive optimization of the IW method, particularly, for $p = 2 + \Omega(1)$, was shown by Braverman and Ostrovsky in [6]. The technique reduces the space requirement of the Hss method by replacing the $\log(mM)$ term by $g(p,n)$, where,

$$g_1(n) = \log n, \ g_c(n) = \log(g_{c-1}(n))/(1 - 2/p), \text{ and } g(p,n) = \min_{c \text{ constant}} g_c(n)$$

Andoni, Krauthgamer and Onak in [2] present an elegant and novel simplification of the IW method. The AKO technique flattens the $O(\log(mM))$ level-wise sampling of the original IW algorithm by replacing each update $(i, v)$ by $(i, w_i^{1/p} \cdot v)$ where, $w_i$'s are drawn from the distribution $f_\mathcal{W}(w) \propto 1/w^2$, for $w \in [1, n^4]$ and are pair-wise independent. Their analysis is based on Preci-sion Sampling [2]. The space used is $O(n^{1-2/p}\epsilon^{-2-4/p}\log(n)p^2 E(p,n))$ words [1]. Here $E(p,n) = (p/(p-2))(1 - n^{-4(1-2/p)})$ and is $O(1)$ for $p = 2 + \Omega(1)$, $O(\log(n))$ for $p = 2 + O(1/\log(n))$ and takes intermediate values in the remainder of the range of $p$ (see Figure 2). A comparison of $E(p,n)$ and $g(p,n)$ is given in Figure 2. The IW, Hss and BO algorithms can be viewed as using a discretized version of the distribution $f_\mathcal{Y} = A/y^3$. These algorithms are "uniform" in that the same sketch structure, using different sizes, can be used to estimate $F_p$ for different values of $p > 2$.

|   | $p = 2 + \Omega(1)$ | $p = 2 + O(1/\log n)$ | $p = 2 + \Theta(1/\log^{(d)} n)$ |
| --- | --- | --- | --- |
| $E(p,n)$ | $O(1)$ | $O(\log n)$ | $O(\log^{(d)} n)$ |
| $g(p,n)$ | $O(\log^{(c)} n)$ | $O(\log n)$ | $O((\log^{(d)} n)^2)$ |

Figure 2: Comparison of $E(p,n)$ with $g(p,n)$. Let $\log^{(c)}(n)$ denote the iterated logarithm of $n$ taken $c$ times, $c$ is a constant.

In this paper, we present an algorithm for estimating $F_p$, $p > 2$. The space used is $O(n^{1-2/p}\epsilon^{-2}\log(n)p^2 E(p,n)/\min(\log(n), \epsilon^{4/p-2}))$ words with word size $O(\log(nmM))$ bits. The update time is $O(\log(n))$. Since $E(p,n) = O(g(p,n))$, this algorithm improves on the space usage over previous algorithms by a factor of $\Omega(\epsilon^{-4/p}\min(\log n, \epsilon^{4/p-2})) \geq \Omega(\epsilon^{-4/p})$. For $\epsilon^{4/p-2} = \Omega(\log(n))$, the space requirement is $O(n^{1-2/p}\epsilon^{-4/p}\log(n)E(p,n))$ words, and for $\epsilon^{4/p-2} = O(\log n)$, the space requirement is $O(n^{1-2/p}\epsilon^{-2}E(p,n))$ words. Note that the algorithm is meaningful only for $2 < p < \log(n)$ and $\epsilon \in [n^{-1/p}, 1]$, otherwise, it uses $\Omega(n\log(mM))$ space.

Braverman and Ostrovsky in [6] state that a modification of their algorithm for the regime $\log n = o(\log(mM))$ and constant $\epsilon$ requires space $O(n^{1-2/p}\epsilon^{-2-4/p}\log(n\log(mM)))$ words.

*Remark 1.* For $p = 2 + o(1)$, a domain reduction technique of Kane et.al. [12] may be applied to reduce space and time. Let $N(n,\epsilon) = \min(n, (n^{1-2/p}\epsilon^{-2})^{18})$ and $\ell(n,\epsilon) = \log(N(n,\epsilon))$. The modified space and time expressions for the Hss, AKO, BO and this paper's algorithm are obtained by modifying the corresponding space/time expressions in Figure 1 by (a) replacing every occurrence of $\log n$ by $\ell(n,\epsilon)$, (b) replacing $n$ of $E(p,n)$ and $g(p,n)$ by $N$, and, (c) adding an $O(\log\log n)$ term for space. For example, for $p = 2 + O(1/\log(n))$, applying domain reduction to our algorithm gives a space bound of $O(\epsilon^{-4/p}\log^2(1/\epsilon) + \log\log n)$ and a time bound of $O(\log(1/\epsilon))$.

*Overview.* The Taylor polynomial estimator is designed for estimating $\psi(\mathbb{E}[X])$ where $X$ is a random variable and $\psi$ is a function with certain smoothness properties. For suitably defined heavy items, the estimator has low bias, controlled variance, and uses a small (logarithmic) number of samples for many functions $\psi$. We apply this technique for $F_p$ estimation by letting $\psi(x) = x^p$. The algorithm uses a pair of variants of the COUNTSKETCH structures, denoted HH and TPEst, respectively, for identifying the heavy-hitters, and for estimating their $p$th frequency powers using the Taylor polynomial estimator. These structures have $O(\log n)$ hash tables, with the tables having $O(p^2 B)$ buckets. Corresponding to each stream update of the form $(i, v)$, the structures HH and TPEst are updated by scaling $v$ to $v \cdot y_i$, so that the effective frequency is $g_i = f_i y_i$. The $y_i$'s are pair-wise independent and are drawn from the distribution $\mathcal{Y} = \mathcal{Y}_p$ with density $f_\mathcal{Y}(y) \propto 1/y^{p+1}$, $y \in [1, n^4]$. A set $H_g$ of heavy items is identified as those whose effective frequency crosses a certain threshold $T_g$–their estimates $\hat{g}_i$ for $|g_i|$ have error at most $T_g/(10p)$. It is then observed that for each heavy item $i$, (a) $\hat{f}_i = \hat{g}_i/y_i$ is on expectation $|f_i|$ with error at most $\min(|f_i|/(10p), T_g/(10p))$, and, (b) assuming sufficient independence of the hash functions, there are $\Omega(\log n)$ tables in TPEst where $i$ does not collide with any other heavy item. The AMS sketches in these non-collision buckets give us $\Omega(\log n)$ estimates for $|f_i|$. The Taylor polynomial is applied to this set of estimates to give an estimate $\bar{\vartheta}_i$ for $|f_i|^p$. Conditioning on an event $\mathcal{G}$ that holds with sufficiently high constant probability, $|\mathbb{E}[\bar{\vartheta}_i] - |f_i|^p| \le O(\epsilon^{12}|f_i|^p)$, $\mathsf{Var}[\bar{\vartheta}_i] \le O(|f_i|^{2p-2}T_g^2/(p^2 \log(n)))$. $\bar{\vartheta}_i$ is scaled by the inverse of the probability that the item qualifies as a heavy item and the sum is taken over the heavy items. The latter probability cannot be obtained accurately enough due to estimation errors. This is resolved by introducing a small rejection probability.

---
[1] The dependency on $\epsilon$ is reduced to $\epsilon^{-2-4/p}$ [3].

## 2 Taylor polynomial estimator

Let $X$ be a random variable with $\mathbb{E}[X] = \mu$ and $\mathsf{Var}[X] = \sigma^2$. The problem is to design an estimator $\theta$ for $\psi(\mu)$ such that (1) $|\mathbb{E}[\theta] - \psi(\mu)| \le O(\epsilon \psi(\mu))$ and (2) $\mathsf{Var}[\theta] = O(\epsilon^2 \psi^2(\mu))$, particularly when $\psi(\mu)$ is large. Singh in [16] proposed an unbiased estimator for $\psi(\mu)$ for an analytic function $\psi$, given an estimate $\lambda$ of $\mu$. Let $X_j, j \ge 0$ be independent copies of $X$. Choose $N$ from the geometric distribution, that is, $\Pr[N = j] = 1/2^{j+1}$, $j \in \mathbb{Z}^+ \cup \{0\}$ and define

$$\theta(\psi, \lambda, \{X_j\}) = 2^{N+1}(\psi^{(N)}(\lambda)/N!)(X_1 - \lambda)(X_2 - \lambda)\ldots(X_N - \lambda) \quad [16] \ .$$

Then $\mathbb{E}[\theta(\psi, \lambda)] = \psi(\mathbb{E}[X])$ [16] and for $\eta^2 = \mathbb{E}[(X - \lambda)^2] = (\mu - \lambda)^2 + \sigma^2$,

$$\mathsf{Var}[\theta(\psi, \lambda, \{X_j\})] = [2\psi^2(\lambda) - \psi^2(\mu)] + 2\sum_{j \ge 1}\left(\frac{\psi^{(j)}(\lambda)}{j!}\right)^2 \eta^{2j} \quad [7] \ .$$

However, the variance is still too large compared to $\epsilon^2 \psi^2(\mu)$.

**Taylor polynomial estimator.** Let $X_j$'s be identically and independently distributed with $\mathbb{E}[X_j] = \mu$ and $\mathsf{Var}[X_j] = \sigma^2$. Also let the first $k+1$ derivatives of $\psi$ exist in the region $[\mu, \lambda]$. Define the estimator $\vartheta$ as follows.

$$\vartheta(\psi, \lambda, k) = \vartheta(\psi, \lambda, k, \{X_j\}_{1 \le j \le k}) = \sum_{j=0}^{k} \frac{\psi^{(j)}(\lambda)}{j!} \prod_{l=1}^{j}(X_l - \lambda) \ .$$

Let $\eta^2 = \mathbb{E}[(X_j - \lambda)^2] = \sigma^2 + (\mu - \lambda)^2$.

**Lemma 1** *Let the first $k+1$ derivatives of $\psi$ exist in the region $[\mu, \lambda]$. Then,*

$$|\mathbb{E}[\vartheta] - \psi(\mu)| = \left|\frac{\psi^{(k+1)}(\lambda')}{(k+1)!}(\mu - \lambda)^{k+1}\right|, \ \lambda' \in (\mu, \lambda), \ and \ \mathsf{Var}[\vartheta] \le \Big(\sum_{1 \le j \le k} \Big|\frac{\psi^{(j)}(\lambda)}{j!}\Big|\eta^j\Big)^2 \ .$$

**Proof** By independence of $X_l$'s,

$$\mathbb{E}[\vartheta] = \sum_{j=0}^{k} \frac{\psi^{(j)}(\lambda)}{j!}(\mu - \lambda)^j = \psi(\mu) - \frac{\psi^{(k+1)}(\lambda')}{(k+1)!}(\mu - \lambda)^{k+1}$$

for some $\lambda' \in (\mu, \lambda)$, implying the first statement of the lemma.

$$\text{Var}[\vartheta] = \mathbb{E}\Big[\Big(\sum_{j=0}^{k} \frac{\psi^{(j)}(\lambda)}{j!} \prod_{l=1}^{j}(X_l - \lambda)\Big)^2\Big] - \Big(\sum_{0 \leq j \leq k} \frac{\psi^{(j)}(\lambda)}{j!}(\mu - \lambda)^j\Big)^2$$

$$= \sum_{j=1}^{k} \Big(\frac{\psi^{(j)}(\lambda)}{j!}\Big)^2 \eta^{2j} + 2 \sum_{1 \leq j < j' \leq k} \frac{\psi^{(j)}(\lambda)}{j!} \cdot \frac{\psi^{(j')}(\lambda)}{j'!} \eta^{2j}(\mu - \lambda)^{j'-j}$$

$$- \Big(\sum_{1 \leq j \leq k} \frac{\psi^{(j)}(\lambda)}{j!}(\mu - \lambda)^j\Big)^2 \leq \Big(\sum_{1 \leq j \leq k} \Big|\frac{\psi^{(j)}(\lambda)}{j!}\Big| \eta^j\Big)^2$$

where, in the second step, the $2k+1$ terms corresponding to $j = 0$ cancel and in the third step we use $\eta \geq |\mu - \lambda|$. ∎

For $p > 0$ and real and $j$ non-negative integral, define $\binom{p}{j} = p(p-1)(p-2)\ldots(p-j+1)/j!$. Then, $\frac{d^j}{dx^j} x^p = \binom{p}{j} x^{p-j}$. Corollary 2 applies the Taylor polynomial estimator to $\psi(x) = x^p$. The proof is given in Appendix A.

**Corollary 2** *Let $\psi(x) = x^p$, $p > 2$, $||\hat{f}_i| - |f_i|| \leq \sigma$, $|f_i| > 9p\sigma$, $k \geq 4\lceil \log(1/\epsilon) \rceil + 8$ and denote $\vartheta(x^p, \hat{f}_i, k)$ by $\vartheta_i$. Then, if $p$ is integral and $k + 1 \geq p$ then the bias is 0. Otherwise, $\big|\mathbb{E}[\vartheta_i] - |f_i|^p\big| \leq |f_i|^p 12^{-8} \epsilon^{12}$. Further, $\text{Var}[\vartheta_i] \leq 3p^2 |f_i|^{2p-2} \sigma^2$.*

**Averaged Taylor polynomial estimator.** Let $\{X_l\}_{l=1}^s$ be a family of independent and identical estimators with expectation $\mu$ and variance $\sigma^2$ and let $\lambda$ be an estimate of $\mu$. Let $s \geq 16k$ and $r = \Theta(s)$. Choose independently $r$ random permutations over $[s]$, denoted $\pi_1, \ldots, \pi_r$. For each permutation $\pi_j$, choose a random permutation $\pi'_j$ of the set $\pi_j([k]) = \{\pi_j(1), \ldots, \pi_j(k)\}$. Let $\tau_j$ denote $\pi'_j \circ \pi_j$. Order the elements of $\tau_j([k])$ in increasing order of the indices, that is, let $\tau_j([k]) = \{a_{j1}, a_{j2}, \ldots, a_{jk}\}$ where $a_{j1} < a_{j2} < \ldots < a_{jk}$. The averaged Taylor polynomial estimator $\bar{\vartheta}$ is defined as follows:

$$\vartheta_j = \sum_{v=0}^{k} \frac{\psi^{(v)}(\lambda)}{v!} \prod_{l=1}^{v}(X_{a_{jl}} - \lambda), \qquad \bar{\vartheta}(\psi, \lambda, k, r, s, \{X_l\}_{l=1}^s) = (1/r) \sum_{j=1}^{r} \vartheta_j \ .$$

The averaged polynomial estimator is also abbreviated as $\bar{\vartheta}(\psi, \lambda, k, r, s)$.

**Lemma 3** $\mathbb{E}\big[\bar{\vartheta}(\psi, \lambda, k, r, s)\big] = \mathbb{E}[\vartheta(\psi, \lambda, k)]$. *For $\psi(x) = x^p, \lambda = \hat{f}_i, \mu = |f_i|, ||\hat{f}_i| - |f_i|| \leq \sigma, |f_i| \geq 9p\sigma$, $k = 144$, $s = 16k$ and $r = 12s$. Then,*

$$\text{Var}\big[\bar{\vartheta}(x^p, \hat{f}_i, k, r, s)\big] \leq (1.5p^2/s)|f_i|^{2p-2} \sigma^2 \ .$$

The proof of Lemma 3 is given in Appendix A.

## 3 Algorithm

The algorithm uses $\text{HH}(C, s)$ and $\text{TPEst}(C, s)$ structures, where, $s = 32 \max(4\lceil \log n \rceil + 4, 144)$,

$$B = \frac{1000 n^{1-2/p} \mathbb{E}_{\mathcal{Y}}[y^2]}{\epsilon^2 \min(\log(n), \epsilon^{4/p-2})}, C = 121 p^2 B, R = n^{4/p} \ \mathbb{E}_{\mathcal{Y}}[y^2] = \frac{p(1 - R^{-(p-2)})}{(p-2)(1 - R^{-p})},$$

where, the distribution $\mathcal{Y}$ is defined below. The $\mathsf{HH}(C,s)$ structure is a COUNTSKETCH$(C,s)$ structure, that is, there are $s$ hash tables $T_1, \ldots, T_s$ each consisting of $C$ buckets. The $\mathsf{TPEst}(C,s)$ structure is a COUNTSKETCH$(C,s)$ structure except that, (a) the hash functions $h_l$'s used for the hash tables $T_l$'s are 3-wise independent, and, (b) the Rademacher family $\{\xi_l(i)\}_{i \in [n]}$ is 4-wise independent for each $l$ and is independent across the $l$'s. Corresponding to each stream update $(i, v)$, the $\mathsf{HH}$ and $\mathsf{TPEst}$ structures are maintained as if the update was $(i, v \cdot y_i)$. The $y_i$'s are pair-wise independent random variables that are chosen from a (discretized version of the) distribution $\mathcal{Y}$ whose density function is given by $f_\mathcal{Y}(x) = x^{-(p+1)}$, $x \in [1, R]$, where, $A = A_p = p(1 - R^{-p})^{-1}$ is the normalization constant. We will show that discretizing the distribution with precision $O(\log(nmM))$ bits is sufficient. The effective frequency of an item is given by $g_i = f_i y_i$. $\mathsf{HH}$ is used to obtain an approximate heavy-hitter set $(H_g, \{\hat{g}_i, \widetilde{\operatorname{sgn}}(g_i)\}_{i \in H_g})$ with threshold $T_g$ and error $\Delta_g$, where, $H_g \subset [n]$, $\hat{g}_i$ and $\widetilde{\operatorname{sgn}}(g_i)$ are estimates for $|g_i|$ and $\operatorname{sgn}(g_i)$ respectively, such that with probability $63/64$, all the following conditions hold: (1) if $|g_i| \geq T_g$ then $i \in H_g$, (2) if $|g_i| < T_g - \Delta_g$, then, $i \notin H_g$, (3) $|\hat{g}_i - |f_i|| \leq \Delta_g$, and (4) $\widetilde{\operatorname{sgn}}(g_i) = \operatorname{sgn}(g_i)$. The parameters are:

$$T_g = \left(\frac{16\mathbb{E}_\mathcal{Y}[y^2] F_2}{B}\right)^{1/2}, \Delta_g = T_g/(11p) \text{ and } \mathbb{E}_\mathcal{Y}[y^2] = \frac{A(1 - R^{-(p-2)})}{p - 2} \qquad (1)$$

For $i \in H_g$, define the estimate $\hat{f}_i$ of $|f_i|$ as $\hat{g}_i/y_i$. Let $l(\hat{f}_i) = \lceil 2\log(2\hat{T}_g/\hat{f}_i)\rceil$, where, $\hat{T}_g \in [T_g, 65T_g/64]$. Let $H = \{i \in H_g : y_i \geq 2^{l(\hat{f}_i)/2}\}$.

Define $\mathrm{NC}(H_g)$ to hold if for every $i \in H_g$, there is a set of at least $s/2$ distinct table indices $Q(i) \subset [s]$ such that $i$ does not collide with any other item of $H_g$ in the buckets to which $i$ maps for the table indices in $Q(i)$. That is,

$$\mathrm{NC}(H_g) \equiv \forall i \in H_g, \exists Q(i) \subset [s], |Q(i)| \geq s/2 \big[\forall j \in H_g \setminus \{i\}, \forall t \in Q(i), \ h_t(i) \neq h_t(j)\big] \ .$$

The estimator $\Theta$ is as follows. If $\mathrm{NC}(H_g)$ does not hold then $\Theta = 0$ (i.e., it fails). Otherwise, let $k = \max(4\lceil \log n \rceil + 4, 144)$ and $r = 12s$. Let $\hat{F}_2 \in (1 \pm \frac{1}{64})F_2$.

$$\nu_{il} = T_l[h_l(i)] \cdot \xi_l(i) \cdot \widetilde{\operatorname{sgn}}(g_i)/y_i, \quad l \in Q(i), \Lambda_i = \{\nu_{il}\}_{l \in Q(i)},$$
$$\bar{\vartheta}_i = \bar{\vartheta}(x^p, \hat{f}_i, k, r, s/2, \Lambda_i),$$
$$\Theta = \sum \left\{\frac{\bar{\vartheta}_i}{\Pr_\mathcal{Y}[y \geq 2^{l(\hat{f}_i)/2}]} \mid i \in H \text{ and } \hat{f}_i > (\epsilon^{2/p} \hat{F}_2/4n)^{1/2}\right\} \ .$$

**Analysis**

We note that the analysis is meaningful for $\epsilon$ in the range $[n^{-1/p}, 1]$. For $\epsilon = o(n^{-1/p})$, the algorithm uses space $\Omega(n \log(mM))$, making it possible for a trivial algorithm to store the entire stream and return $F_p$ exactly.

Let $G_2 = \sum_{i \in [n]} g_i^2$. Since, $\mathbb{E}[G_2] = F_2 \mathbb{E}_\mathcal{Y}[y_i^2]$, we have, $G_2 \leq 16 F_2 \mathbb{E}_\mathcal{Y}[y_i^2]$ with probability $15/16$. The analysis is conditioned on the conjunction of the following events, denoted by $\mathcal{G}$.

(1) $\mathcal{G}_{G_2}(B) \equiv G_2 \leq 16 F_2 \mathbb{E}_\mathcal{Y}[y_i^2]$, (2) $\mathcal{G}_H \equiv H_g$ is a $(T_g, \Delta_g)$ approximate heavy set, and (3) $\mathrm{NC}(H_g)$ .

From properties of COUNTSKETCH, $\mathcal{G}_H$ holds with probability $63/64$. Conditional on $\mathcal{G}_H$ and $\mathcal{G}_{G_2}$, $|H_g| \le G_2/(T_g - \Delta_g)^2 \le B(1-B/C)^{-1} \le B(1+1/(10p)^2)$. Since the hash functions of the TPEst structure are 3-wise independent hence, $\mathrm{NC}(H_g)$ holds with probability $1 - n^{-10}$. (see Appendix A.) Hence $\mathcal{G}$ holds with probability at least $1 - 1/64 - 1/16 - n^{-10} \ge 58/64$.

Let $\xi$ denote the Rademacher random variables in the TPEst structure and let $\zeta_H$ and $\zeta_E$ denote the random bits used by the HH and TPEst structures respectively (so that $\zeta_E$ includes the random seed for $\xi$ and the hash functions of TPEst) and let $\zeta$ refer jointly to all these random bits. Then,

$$\mathbb{E}_\xi\bigl[\nu_{iu} \mid \mathcal{G}, \zeta_H\bigr] = \mathbb{E}_\xi\bigl[y_i^{-1}\bigl(|g_i| + \sum_{j \ne i, h_u(i) = h_u(j)} g_j \xi_u(j) \xi_u(i) \widetilde{\mathrm{sgn}}(g_i)\bigr)\bigr] = \frac{|g_i|}{y_i} = |f_i| \ .$$

Lemma 4 shows that as a consequence of $\mathrm{NC}(H_g)$, for any $i, j$ from $H_g$ and distinct, the expectation of the product of any subset of $\nu_{il}$'s for $l \in Q(i)$ and any subset of $\nu_{jl'}$'s for $l' \in Q_j$ is the product of their expectations.

**Lemma 4** $\mathbb{E}_\xi\bigl[\bar{\vartheta}_i \bar{\vartheta}_j \mid \mathcal{G}, \zeta_H\bigr] = \mathbb{E}_\xi\bigl[\bar{\vartheta}_i \mid \mathcal{G}, \zeta_H\bigr] \mathbb{E}_\xi\bigl[\bar{\vartheta}_j \mid \mathcal{G}, \zeta_H\bigr]$, where, $\zeta_H$ is any choice of random bits such that $i, j \in H$.

**Proof** We first show that if $i, j \in H_g$, $i \ne j$, $A_i \subset Q(i)$ and $A_j \subset Q(j)$, then,
$\mathbb{E}_\xi\bigl[\prod_{u \in A_i, u' \in A_j} \nu_{iu} \nu_{ju} \mid \mathcal{G}, \zeta_H\bigr] = |f_i|^{|A_i|} |f_j|^{|A_j|}$.

$\nu_{iu}$ and $\nu_{ju'}$ are independent if $u \ne u'$ since the inference is made from different tables. Also they are independent of $\zeta_H$. Hence,

$$\mathbb{E}_\xi\Bigl[\prod_{u \in A_i} \nu_{iu} \prod_{u' \in A_j} \nu_{ju'} \mid \mathcal{G}, \zeta_H\Bigr] = |f_i|^{|A_i \setminus A_j|} |f_j|^{|A_j \setminus A_i|} \prod_{u \in A_i \cap A_j} \mathbb{E}_\xi\bigl[\nu_{iu} \nu_{ju} \mid \mathcal{G}\bigr].$$

If $u \in A_i \cap A_j$, then, due to $\mathrm{NC}(H_g)$ and since $i, j \in H \subset H_g$, $h_u(i) \ne h_u(j)$. From 4-wise independence of the $\xi_u$ family, it follows that $\mathbb{E}_\xi\bigl[\nu_{ir} \nu_{jr} \mid \mathcal{G}, \zeta_H\bigr] = |f_i||f_j|$. The above expectation becomes $|f_i|^{|A_i \setminus A_j|} |f_j|^{|A_j \setminus A_i|} (|f_i||f_j|)^{|A_i \cap A_j|} = |f_i|^{|A_i|} |f_j|^{|A_j|}$.

Arguing in an identical manner, and noting that $\hat{f}_i, \hat{f}_j$ are dependent only on $\zeta_H$ and are independent of $\xi$, we have

$$\mathbb{E}_\xi\Bigl[\prod_{u \in A_i}(\nu_{iu} - \hat{f}_i) \prod_{u' \in A_j}(\nu_{ju} - \hat{f}_j) \mid \mathcal{G}, \zeta_H\Bigr] = \bigl||f_i| - \hat{f}_i\bigr|^{|A_i|} \bigl||f_i| - \hat{f}_i\bigr|^{|A_j|} \quad (2)$$

Let $i, j \in H_g$ and $i \ne j$. Denote $\psi^{(v)}(\lambda)/v!$ by $a_v(\lambda)$. Then, for random permutations $\tau_i, \tau_j$ chosen by the averaged polynomial estimator,

$$\begin{aligned} &\mathbb{E}_\xi\bigl[\vartheta_i \vartheta_j \mid \mathcal{G}, \zeta_H\bigr] \\ &= \sum_{0 \le v, v' \le k} a_v(\hat{f}_i) a_{v'}(\hat{f}_j) \mathbb{E}_\xi\Bigl[\prod_l (\nu_{il} - \hat{f}_i) \prod_{l'} (\nu_{jl'} - \hat{f}_j) \mid \mathcal{G}, \zeta_H\Bigr] \\ &= \sum_{0 \le v, v' \le k} a_v(\hat{f}_i) a_{v'}(\hat{f}_j) \bigl||f_i| - \hat{f}_i\bigr|^v \bigl||f_i| - \hat{f}_i\bigr|^{v'} \\ &= \mathbb{E}_\xi\bigl[\vartheta_i \mid \mathcal{G}, \zeta_H\bigr] \mathbb{E}_\xi\bigl[\vartheta_j \mid \mathcal{G}, \zeta_H\bigr] \end{aligned} \quad (3)$$

The variables $l, l'$ run over the least $v$ indices of $\tau_i([k])$ and the least $v'$ indices of $\tau_j([k])$, respectively. The second step follows from (2). The values of $\hat{f}_i, \hat{f}_j$ are dependent only on $\zeta_H$, and therefore is the same in the corresponding expressions for $\mathbb{E}_\xi[\vartheta_i \mid \mathcal{G}, \zeta_H]$ and $\mathbb{E}_\xi[\vartheta_j \mid \mathcal{G}, \zeta_H]$. Hence, by (3),

$$\mathbb{E}_\xi[\bar{\vartheta}_i \bar{\vartheta}_j \mid \mathcal{G}, \zeta_H] = \frac{1}{r^2} \sum_{1 \le t,t' \le r} \mathbb{E}_\xi[\vartheta_i(S_{it}, \tau_{it})\vartheta_j(S_{jt'}, \tau_{jt'}) \mid \mathcal{G}, \zeta_H]$$
$$= \mathbb{E}_\xi[\bar{\vartheta}_i \mid \mathcal{G}, \zeta_H]\mathbb{E}_\xi[\bar{\vartheta}_j \mid \mathcal{G}, \zeta_H] \;. \blacksquare$$

Lemma 5 shows that for $i \in H$, $\hat{f}_i$ is an accurate estimate for $|f_i|$.

**Lemma 5** *Given $\mathcal{G}$ and $i \in H$, $|\hat{f}_i - |f_i|| \le (1/(10p))|f_i|$.*

**Proof** Since $i \in H$, $|f_i|y_i = |g_i| \ge T_g - \Delta_g$ or, $y_i \ge \max(1, (T_g - \Delta_g)/|f_i|)$. Also,

$$\Delta_g \ge |\hat{g}_i - g_i| = |\hat{f}_i - |f_i||y_i \ge |\hat{f}_i - |f_i||(T_g - \Delta_g)/|f_i|$$
$$\text{or,} \quad |\hat{f}_i - |f_i|| \le \frac{\Delta_g}{T_g - \Delta_g}|f_i| \le (1 - (B/C)^{1/2})^{-1}(B/C)^{1/2}|f_i| \le \frac{|f_i|}{10p} \;.$$

since, $\Delta_g/T_g \le (B/C)^{1/2} = 1/(11p)$. $\blacksquare$

**Lemma 6** *Given $\zeta_H$ such that $i \in H$, $\mathsf{Var}_{\zeta_E}[\nu_{il} \mid \mathcal{G}, \zeta_H] \le \min(\Delta_g^2, \frac{f_i^2}{(10p)^2})$.*

**Proof** Since $i \in H$, $y_i \ge \max(1, (T_g - \Delta_g)/|f_i|)$. Since $\mathsf{TPEst}(C)$ is a COUNTSKETCH structure and since $\mathcal{G}$ holds, $G_2 \le T_g^2 B = \Delta_g^2 C$, and $\mathrm{sgn}(g_i) = \widetilde{\mathrm{sgn}}(g_i)$. So

$$\mathsf{Var}_{\zeta_E}[\nu_{il} \mid \mathcal{G}, \zeta_H] = y_i^{-2}\mathsf{Var}_{\zeta_E}[T_l[h_l(i)] \cdot \xi_l(i) \cdot \widetilde{\mathrm{sgn}}(g_i) \mid \mathcal{G}, \zeta_H]$$
$$\le \min\left(1, \frac{f_i^2}{(T_g - \Delta_g)^2}\right)\frac{G_2}{C} \le \min\left(\Delta_g^2, \frac{f_i^2 \Delta_g^2}{(T_g - \Delta_g)^2}\right) \le \min\left(\Delta_g^2, \frac{f_i^2}{(10p)^2}\right) \;. \blacksquare$$

Let $x_i$ denote the indicator variable that is 1 if $i \in H$ and 0 otherwise. By Lemma 5, conditional on $\mathcal{G}$, if $i \in H_g$ and $y_i \ge 2^{l(\hat{f}_i)/2}$, then, $i \in H$. Equivalently, $x_i = \omega_i(y_i) = \mathbf{1}_{y_i \ge 2^{l(\hat{f}_i)/2}}$ and $\mathbf{1}_P$ is 1 if $P$ is true and is 0 otherwise.

**Lemma 7** $\mathbb{E}_{\zeta \setminus \xi}[x_i/\mathsf{Pr}_\mathcal{Y}[y \ge 2^{l(\hat{f}_i)/2}] \mid \mathcal{G}] = 1$.

**Proof** We note that $\hat{f}_i$ and $y_i$ depend only on $\zeta \setminus \xi$ and drop the suffix from the expectation. Conditional on $\mathcal{G}$, the predicate $y_i \ge 2^{l(\hat{f}_i)/2}$ is equivalent to $i \in H$. Actually, the condition for $i \in H$ is (1) $i \in H_g$ and (2) $y_i \ge 2^{l(\hat{f}_i)/2}$. However, conditional on $\mathcal{G}$, the second condition implies the first and hence the equivalence. Further, if $i \in H$, then $\hat{f}_i \in |f_i|(1 \pm c)$, where, $c = \frac{1}{10p}$ by Lemma 5.

*Simple Case*: $l(s)$ is constant in the interval $s \in [|f_i|(1-c), |f_i|(1+c)]$. Then,

$$\mathbb{E}\left[\frac{x_i}{\mathsf{Pr}_\mathcal{Y}[y \ge 2^{l(\hat{f}_i)/2}]} \mid \mathcal{G}\right] = \frac{1}{y \ge 2^{l(|f_i|)/2}}\mathsf{Pr}[x_i = 1 \mid \mathcal{G}] = 1 \;.$$

*General Case*: $l(s)$ is not constant in the interval $[|f_i|(1-c), |f_i|(1+c)]$. Let $\phi(s,y)$ denote the probability density function that $y_i = y$ and $\hat{f}_i = s$, conditional on $\mathcal{G}$. Such a function $\phi$ exists since there is a non-trivial probability that $i \in H$ conditional on $\mathcal{G}$ for each $i \in [n]$. So

$$\mathbb{E}\left[\frac{x_i}{\Pr_{\mathcal{Y}}[y \geq 2^{l(\hat{f}_i)/2}]} \mid \mathcal{G}\right] = \int_{s,y} \frac{\phi(y,s)\omega_i(y)}{\Pr_{\mathcal{Y}}[y' \geq 2^{l(s)/2}]} ds dy \qquad (4)$$

We have

$$l(s) = \left\lceil \log \frac{2\hat{T}_g}{s} \right\rceil \in \left\lceil \log \frac{2\hat{T}_g}{|f_i|(1 \pm c)} \right\rceil \in \left\lceil \log \frac{2\hat{T}_g}{|f_i|} - \log(1 \pm c) \right\rceil \in [l_{1i}, l_{2i}] = \{l_{1i}, l_{1i+1}\}$$

since (a) $c \leq 1/20$ and therefore, $l_{2i} = l_{1i} + 1$ and (b) $l(s)$ is an integer.

If $sy < 2\hat{T}_g$, then, $y < 2\hat{T}_g/s < 2^{l(s)/2}$ and so $\phi(y,s) = 0$. Hence, the region of integration is restricted to $R = \{(s,y) : sy \geq 2\hat{T}_g\}$. Sub-divide $R$ into (1) $R_1 : (s,y) \in R$ and $l(s) = l_{1i}$, and, (2) $R_2 : (s,y) \in R$ and $l(s) = l_{2i}$. Let $\phi_1(s,y)$ denote the joint probability density of $s, y$ given that $(s,y) \in R_1$, conditional on $\mathcal{G}$. Likewise, define $\phi_2(s,y)$. Let $p = \Pr[(s,y) \in R_1 \mid \mathcal{G}]$ and let $q = 1 - p$. From (4), we have,

$$\mathbb{E}\left[\frac{x_i}{\Pr_{\mathcal{Y}}[y \geq 2^{l(\hat{f}_i)/2}]} \mid \mathcal{G}\right] = \int_{(s,y) \in R} \frac{\phi(s,y)\omega_i(y) ds dy}{\Pr_{\mathcal{Y}}[y' \geq 2^{l(\hat{f}_i)/2}]}$$
$$= p \int_{(s,y) \in R_1} \frac{\phi_1(s,y)\omega_i(y) ds dy}{\Pr_{\mathcal{Y}}[y' \geq 2^{l_{1i}/2}]} + q \int_{(s,y) \in R_2} \frac{\phi_2(s,y)\omega_i(y) ds dy}{\Pr_{\mathcal{Y}}[y' \geq 2^{l_{2i}/2}]}$$
$$= \frac{p \cdot \Pr[x_i = 1 \mid l(\hat{f}_i) = l_{1i}]}{\Pr_{\mathcal{Y}}[y' \geq 2^{l_{1i}/2}]} + \frac{q \cdot \Pr[x_i = 1 \mid l(\hat{f}_i) = l_{2i}]}{\Pr_{\mathcal{Y}}[y' \geq 2^{l_{2i}/2}]} = p + q = 1 \; . \quad \blacksquare$$

Let $\rho_i(\hat{f}_i) = \Pr_{\mathcal{Y}}[y_i \geq 2^{l(\hat{f}_i)/2}]$.

**Lemma 8** *For $i, j \in [n]$ and distinct, $\mathbb{E}_{\zeta \setminus \xi}\left[\frac{x_i x_j}{\rho_i \rho_j} \mid \mathcal{G}\right] = 1$.*

**Proof** Suppose that $l(s_i)$ and $l(s_j)$ are respectively constant for $s_i \in |f_i|(1 \pm c)$ and $s_j \in |f_j|(1 \pm c)$ (i.e., the simple case of the previous proof).

$$\mathbb{E}\left[\frac{x_i x_j}{\rho_i(\hat{f}_i)\rho_j(\hat{f}_j)} \mid \mathcal{G}\right] = \frac{\Pr[y_i \geq 2^{l(|f_i|)/2}, y_j \geq 2^{l(|f_j|)/2}]}{\rho_i(|f_i|)\rho_j(|f_j|)}$$
$$= \frac{\Pr[y_i \geq 2^{l(|f_i|/2)}]\Pr[y_j \geq 2^{l(|f_j|/2)}]}{\rho_i(|f_i|)\rho_j(|f_j|)} = \frac{\rho_i(|f_i|)\rho_j(|f_j|)}{\rho_i(|f_i|)\rho_j(|f_j|)} = 1$$

where step 2 follows by the pair-wise independence of $y_i$ and $y_j$. The other cases proceed analogously. $\blacksquare$

Assume $\mathcal{G}$ and let $\psi(x) = x^p$. Suppose $i \in H$ holds for the given $\zeta_H$. Then,

(1) $\mathbb{E}_\xi[\nu_{il} \mid \mathcal{G}, \zeta_H] = |f_i|, l \in Q(i)$ (2) $|g_i| \geq T_g - \Delta_g \geq T_g(1 - 1/(11p))$
(3) $|\hat{f}_i - |f_i|| \leq |f_i|/(10p)$, and (4) $\mathsf{Var}_\xi[\nu_{il} \mid \mathcal{G}, \zeta_H] \leq \min(\Delta_g^2, |f_i|^2/(10p)^2)$

Hence, all the premises of Corollary 2 and Lemma 3 are satisfied, with $\sigma = \mathsf{Var}_\xi[\nu_{il} \mid \mathcal{G}, \zeta_H]$. Hence, by Corollary 2, $\mathbb{E}_\xi[\bar{\vartheta}_i \mid \mathcal{G}, \zeta_H] \in |f_i|^p(1 \pm 12^{-8}\epsilon^{12})$. Also, $x_i, \rho_i(\hat{f}_i)$ depend only on $\zeta \setminus \xi$ and the seed for $\zeta_H$ is a substring of the seed for $\zeta_L \setminus \xi$. Hence,

$$\mathbb{E}[\bar{\vartheta}_i x_i/\rho_i \mid \mathcal{G}] = \mathbb{E}_{\zeta \setminus \xi}[(x_i/\rho_i)\mathbb{E}_\xi[\bar{\vartheta}_i \mid \mathcal{G}, \zeta \setminus \xi]]$$
$$\in |f_i|^p(1 \pm 12^{-8}\epsilon^{12})\mathbb{E}_{\zeta \setminus \xi}[x_i/\rho_i] \in |f_i|^p(1 \pm 12^{-8}\epsilon^{12}) \quad (5)$$

since, $\mathbb{E}[x_i/\rho_i \mid \mathcal{G}] = 1$ by Lemma 7.

Let

$$L = \{i \in H \mid \hat{f}_i < (\epsilon^{2/p}\hat{F}_2)^{1/2}/(2n^{1/2})\}$$
$$L_1 = \{i \in H \mid |f_i| < (\epsilon^{2/p}F_2/n)^{1/2}\}.$$

By definition, $\Theta = \sum\{\bar{\vartheta}_i/\rho_i(\hat{f}_i) \mid i \in L\}$. Since $i \in H$, $\hat{f}_i \in |f_i|(1 \pm \frac{1}{(10p)})$ and $\hat{F}_2 \in (1 \pm \frac{1}{64})F_2$, hence $L \subset L_1$.

**Lemma 9** *For $n > 2000$, $|\mathbb{E}[\Theta] - F_p \mid \mathcal{G}| \leq \epsilon F_p/16$.*

**Proof** $\sum_{i \in L_1}|f_i|^p \leq n((\epsilon^{2/p}F_2)^{1/2}/n^{1/2})^p \leq \epsilon F_p/n$, since, $F_2 \leq F_p^{2/p} n^{1-2/p}$. So,

$$\left|\mathbb{E}[\Theta \mid \mathcal{G}] - F_p\right| \leq \left|\mathbb{E}\left[\sum_{i \in [n] \setminus L} \bar{\vartheta}_i x_i/\rho_i\right] - F_p\right| \leq \sum_{i \in [n] \setminus L_1} 12^{-8}\epsilon^{12}|f_i|^p + \sum_{i \in L_1}|f_i|^p$$
$$\leq F_p(12^{-8}\epsilon^{12} + \epsilon/n) \leq (\epsilon/1000)F_p \quad (6)$$

where, the second step follows from (5) and triangle inequality for $n > 2000$. ∎

Let $\rho_i^{\min}$ be the smallest value of $\rho_i(\hat{f}_i) = \mathsf{Pr}_\mathcal{Y}[y \geq 2^{l(\hat{f}_i)/2} \mid i \in H]$. Let $c = \frac{1}{10p}$, $H_1 = \{i \mid |f_i| \geq 65T_g/(64(1-c))\}$, $\alpha = \frac{\sqrt{2}(1+1/64)}{1-c}$ and $\omega = \mathbb{E}_\mathcal{Y}[y^2]$.

**Lemma 10** *Conditional on $\mathcal{G}$, if $i \in H_1$, then, $\rho_i^{\min} = 1$ and $x_i = 1$. Otherwise, $\exists$ constant $K_0$ such that for $n > K_0$, $i \in [n] \setminus (H_1 \cup L_1)$ and $\epsilon \geq n^{-1/p}$, $\rho_i^{\min} \geq (1.01)\left(\frac{\alpha T_g}{|f_i|}\right)^{-p}$.*

**Proof** $\rho_i^{\min} = 1$ if $2^{l(\hat{f}_i)/2} \leq 1$ or if, $\frac{T_g(1+1/64)}{|f_i|(1-c)} \leq 1$, that is, if $i \in H_1$. So for $i \in H_1$, $\rho_i = x_i = 1$ (cond. on $\mathcal{G}$). For $i \in [n] \setminus (H_1 \cup L_1)$,

$$\rho_i^{\min} \geq \mathsf{Pr}\left[y \geq 2^{\frac{1}{2}\lceil 2 \log \frac{T_g}{|f_i|(1-c)}\rceil}\right] \geq \mathsf{Pr}\left[y \geq \frac{\alpha T_g}{|f_i|}\right] = (A/p)\left[\left(\frac{\alpha T_g}{|f_i|}\right)^{-p} - n^{-4}\log^{-p}(n)\right]$$

For $i \in [n] \setminus L_1$, $|f_i| \geq (\epsilon^{2/p}F_2/n)^{1/2})$. Since, $T_g = (16\omega F_2/B)^{1/2} \geq \left(\frac{\epsilon^2 F_2 \min(\log(n), \epsilon^{4/p-2})}{64n^{1-2/p}\log(n)}\right)^{1/2}$ and $\epsilon \geq n^{-1/p}$, we have,

$$\left(\frac{\alpha T_g}{|f_i|}\right)^{-p} \geq \left(\frac{\epsilon^{2/p}F_2}{n} \cdot \frac{Kn^{1-2/p}\log(n)}{\epsilon^2 F_2 \min(\log(n), \epsilon^{4/p-2})}\right)^{p/2} \geq K' n^{-2/p-2/p^2} \gg n^{-4} . \quad (7)$$

where, $K, K'$ are constants. So, $\exists$ constant $K_0$ such that for $n > K_0$, $\left(\frac{\alpha T_g}{|f_i|}\right)^{-p} - n^{-4p} \geq (0.999)\left(\frac{\alpha T_g}{|f_i|}\right)^{-p}$. Also, $A = p(1+3n^{-4p})$ and so $(A/p) \leq 1.001$. Therefore, $\rho_i^{\min} \geq (1.01)\left(\frac{\alpha T_g}{|f_i|}\right)^{-p}$. ∎

We have for $i \neq j$,

$$\mathbb{E}_\xi\bigl[\bar{\vartheta}_i\bar{\vartheta}_j x_i x_j/(\rho_i\rho_j) \mid \mathcal{G}\bigr] - \mathbb{E}_\xi\bigl[\bar{\vartheta}_i x_i/\rho_i \mid \mathcal{G}\bigr]\mathbb{E}_\xi\bigl[\bar{\vartheta}_j x_j/\rho_j \mid \mathcal{G}\bigr]$$
$$= \frac{x_i x_j}{\rho_i \rho_j}\Bigl[\mathbb{E}_\xi\bigl[\bar{\vartheta}_i\bar{\vartheta}_j \mid \mathcal{G}, \zeta_H\bigr] - \mathbb{E}_\xi\bigl[\bar{\vartheta}_i \mid G, \zeta_H\bigr]\mathbb{E}_\xi\bigl[\bar{\vartheta}_j \mid \mathcal{G}, \zeta_H\bigr]\Bigr] = 0 \ .$$

by Lemma 4. Here $\zeta_H$ is a choice of the value of the random string such that $i, j \in H$. Hence,

$$\mathsf{Var}_\xi\Bigl[\sum_{i \in [n]\setminus L_2} \bar{\vartheta}_i x_i/\rho_i \mid \mathcal{G}\Bigr] = \sum_{i \in [n]\setminus L_2} \mathsf{Var}_\xi\bigl[\bar{\vartheta}_i x_i/\rho_i \mid \mathcal{G}\bigr] \ . \tag{8}$$

Let

$$L_2 = \{i \in H \mid |f_i| < (\epsilon^{2/p} F_2/(4n))^{1/2}\} \ .$$

**Lemma 11** *Let $\epsilon \geq n^{-1/p}$. Then, there exists a constant $n_0$ such that for $n > n_0$, $\Pr\bigl[|\Theta - F_p| \leq \epsilon F_p\bigr] \geq 0.8$.*

**Proof** Conditional on $\mathcal{G}$, $\hat{f}_i \in (1 \pm \frac{1}{10p})$ and $\hat{F}_2 \in (1 \pm \frac{1}{64})F_2$, so, $L_2 \subset L$.

$$\mathsf{Var}\bigl[\Theta \mid \mathcal{G}\bigr] \leq \mathsf{Var}\Bigl[\sum_{i \in [n]\setminus L_2} \bar{\vartheta}_i x_i/\rho_i \mid \mathcal{G}\Bigr] =$$
$$\mathbb{E}_{\zeta\setminus\xi}\Bigl[\mathsf{Var}_\xi\Bigl[\sum_{i \in [n]\setminus L_2} \bar{\vartheta}_i x_i/\rho_i \mid \mathcal{G}\Bigr]\Bigr] + \mathsf{Var}_{\zeta\setminus\xi}\Bigl[\mathbb{E}_\xi\Bigl[\sum_{i \in [n]\setminus L_2} \bar{\vartheta}_i x_i/\rho_i \mid \mathcal{G}\Bigr]\Bigr] \tag{9}$$

where, the first step follows by definition and since $L_1 \subset L$, the second step is an identity (law of total variance). In the next equation, let $i$ run over $[n] \setminus L_2$.

$$\mathsf{Var}_\xi\Bigl[\sum_{i \in [n]\setminus L_2} \bar{\vartheta}_i x_i/\rho_i \mid \mathcal{G}\Bigr] = \sum_{i \in [n]\setminus L_2} \mathsf{Var}_\xi\bigl[\bar{\vartheta}_i x_i/\rho_i \mid \mathcal{G}\bigr] = \sum_{i \in [n]\setminus L_2} \frac{x_i}{\rho_i^2}\mathsf{Var}_\xi\bigl[\bar{\vartheta}_i \mid i \in H, \mathcal{G}\bigr]$$
$$\leq \sum_{i \in [n]\setminus L_2} \frac{x_i}{\rho_i^2}(3p^2/s)|f_i|^{2p-2}\min(\Delta_g^2, |f_i|^2/(10p)^2), \quad \text{by Lemma 4.}$$

where, Step 1 follows from (8). Hence,

$$\mathbb{E}_{\zeta\setminus\xi}\Bigl[\mathsf{Var}_\xi\Bigl[\sum_{i \in [n]\setminus L_2} \bar{\vartheta}_i x_i/\rho_i \mid \mathcal{G}\Bigr]\Bigr]$$
$$\leq \sum_{i \in [n]\setminus L_2} (\rho_i^{\min})^{-1}(3p^2/s)|f_i|^{2p-2}\min(\Delta_g^2, |f_i|^2/(10p)^2)$$
$$\leq \sum_{i \in H_1} |f_i|^{2p-2}p^2\Delta_g^2/(42\log n) + \sum_{i \in [n]\setminus(H_1 \cup L_2)} |f_i|^p(\alpha T_g)^p/(4200\log n)$$
$$\leq F_{2p-2}\Delta_g^2/(42\log n) + (\alpha T_g)^p F_p/(4200\log n)$$
$$\leq F_p^2((\epsilon/500)^2 + (\epsilon/32)^2/\log(n)) \ . \tag{10}$$

Step 1 follows since, $\mathbb{E}_{\zeta\setminus\xi}[x_i/\rho_i^2] \leq (1/\rho_i^{\min})\mathbb{E}[x_i/\rho_i] = (1/\rho_i^{\min})$. Step 2 uses (a) Lemma 10 which holds for $n > K_0$, (b) $s \geq 128\log n$, and, (c) $\rho_i^{\min} = 1$ for $i \in H_1$. Step 4 uses the following bounds. We have, $F_{2p-2} \leq F_p^{2-2/p}$ since $p \geq 2$ and $F_2 \leq F_p^{2/p} n^{1-2/p}$. From the latter, we have, $F_2^{p/2} \leq F_p (n^{1-2/p})^{p/2}$, Since, $B > 1000\omega\epsilon^{-2} n^{1-2/p}/\min(\log n, \epsilon^{4/p-2})$, we have,

$$F_p(\alpha T_g)^p = F_p(16\alpha\omega F_2/B)^{p/2} \leq (64)^{-p/2} F_p^2 \min(\epsilon^p \log^{p/2}(n), \epsilon^2) \leq (\epsilon/8)^2 F_p^2$$

and

$$F_{2p-2} p^2 \Delta_g^2 \leq F_p^{2-2/p}(16p^2\omega F_2)/((11p)^2 B) \leq (88)^{-2} \epsilon^2 F_p^2 \log n \ .$$

Proceeding analogously for the second summation term in (9),

$$\begin{aligned}
\mathsf{Var}_{\zeta\setminus\xi}\Big[\sum_{i\in[n]\setminus L_1} \mathbb{E}_\xi\big[\bar{\vartheta}_i x_i/\rho_i\big]\Big] \\
= \mathsf{Var}_{\zeta\setminus\xi}\Big[\sum_{i\in[n]\setminus L_1} (x_i/\rho_i)\mathbb{E}_\xi\big[\bar{\vartheta}_i \mid i \in H\big]\Big] \\
\leq \mathsf{Var}_{\zeta\setminus\xi}\Big[\sum_{i\in[n]\setminus L_1} (x_i/\rho_i)|f_i|^p(1 + 12^{-8}\epsilon^{12})\Big] \\
\leq \sum_{i\in[n]\setminus L_1} (1.001)|f_i|^{2p} \mathsf{Var}_{\zeta\setminus\xi}\big[(x_i/\rho_i)\big] \\
\leq \sum_{i\in[n]\setminus(H_1\cup L_1)} 1.001 |f_i|^{2p} (\rho_i^{\min})^{-1} \\
\leq \sum_i 1.001 |f_i|^p (\alpha T_g)^p \leq 1.001 F_p(\alpha T_g)^p \\
\leq (\epsilon/8)^2 F_p^2
\end{aligned} \quad (11)$$

Step 1 follows since $x_i, \rho_i$ are independent of $\xi$, and since $x_i$ is boolean, Step 2 follows from Corollary 2. Step 3 is inferred as follows.

$$\begin{aligned}
\mathsf{Var}_{\zeta\setminus\xi}\Big[\sum_{i\in[n]\setminus L_1} (x_i/\rho_i)|f_i|^p(1+12^{-8}\epsilon^{12})\Big] &= \sum_{i\in[n]\setminus L_1} (|f_i|^p(1+12^{-8}\epsilon^{12}))^2 \mathsf{Var}_{\zeta\setminus\xi}\big[x_i/\rho_i\big] \\
&\quad + 2\sum_{i\neq j} |f_i|^p |f_j|^p (1+12^{-8}\epsilon^{12})^2 \big[\mathbb{E}_{\zeta\setminus\xi}[x_i x_j/(\rho_i\rho_j)] - \mathbb{E}_{\zeta\setminus\xi}[x_i/\rho_i]\mathbb{E}_{\zeta\setminus\xi}[x_i/\rho_i]\big] \\
&= \sum_{i\in[n]\setminus L_1} (|f_i|^p(1+12^{-8}\epsilon^{12}))^2 \mathsf{Var}_{\zeta\setminus\xi}\big[x_i/\rho_i\big] \\
&\leq \sum_{i\in[n]\setminus L_1} 1.001 |f_i|^{2p} \mathsf{Var}_{\zeta\setminus\xi}\big[x_i/\rho_i\big] \ .
\end{aligned}$$

where, the cross term vanishes by Lemma 8. Step 4 in the sequence of steps leading to (11) uses $\mathsf{Var}_{\zeta\setminus\xi}[x_i/\rho_i] \leq \mathbb{E}_{\zeta\setminus\xi}[x_i/\rho_i^2] \leq \rho_i^{\min}\mathbb{E}_{\zeta\setminus\xi}[x_i/\rho_i] = \rho_i^{\min}$. Step 5 uses Lemma 10, and Step 6 is a direct simplification and the final step uses the calculation for $F_p(\alpha T_g)^p$ done earlier in this proof.

Adding (10) and (11), $\mathsf{Var}[\Theta] \leq (\epsilon F_p/7.5)^2$. Hence, for $n > n_0 = \max(K_0, 2000)$, by Chebychev's inequality,

$$\Pr[|\Theta - \mathbb{E}[\Theta]| \leq (\epsilon/1.01) F_p \mid \mathcal{G}] \geq 1/55 \ .$$

By (6), $|\mathbb{E}[\Theta \mid \mathcal{G}] - F_p| \le (\epsilon/1000)F_p$ for $n > 2000$. Hence by triangle inequality,

$$|\Theta - F_p| \le \epsilon F_p \text{ with probability } 1 - 1/55, n > n_0 \ .$$

The above probabilities were conditioned on $\mathcal{G}$, which holds with probability $57/64$. So $|\Theta - F_p| \le \epsilon F_p$ holds unconditionally with probability at least $(54/55)(57/64) \ge 0.8$. ∎

*Discretizing $\mathcal{Y}$.* We note that $p = o(\log(n))$, otherwise, the space required is $\Omega(n \log(mM))$ and a trivial linear space structure can be kept. Since $|f'_{\mathcal{Y}}(x)| \in [R^{-(p+2)}, 1] = [n^{-1-2/p}, 1]$ and $R = n^{4/p}$, it suffices to have a precision of $O(\log(pn)) = O(\log(n))$ bits each before and after the binary point. The resulting word size of each entry of HH and TPEst is $O(\log(nmM))$ bits. We have basically proved the main result of the paper.

**Theorem 12** *For each $p > 2$ and $0 \le \epsilon \le 1$, there exists a sketch structure that can be updated over data streams and on query can return $\hat{F}_p$ satisfying $|\hat{F}_p - F_p| \le \epsilon F_p$ with probability 0.80. The space used in bits is*

$$\min\left(n\log(mM), \frac{(p/(p-2))(1 - n^{-4(1-2/p)})p^2 n^{1-2/p}(\log n)\log(nmM)}{\epsilon^2 \min(\log(n), \epsilon^{4/q-2})}\right) \ .$$

*The update time is $O((\log n))$.*

**Proof** If $\epsilon < n^{1/p}$ or $p \ge \log(n)$, we store the stream trivially using $O(n \log(mM))$ bits and return $F_p$. Otherwise, the space requirement follows from Lemma 11 and discretization argument above. The update time is dominated by $s = O(\log n)$ hash function evaluations. The hash functions are pair-wise independent for the $\mathsf{HH}(C, s)$ structure, 3-wise independent for $\mathsf{TPEst}(C, s)$ structure and the Rademacher sketches require pair-wise and 4-wise independence for the HH and TPEst structures, respectively. Thus, the update time is $O(\log n)$. ∎

## A    Proofs and Notes

*A slightly modified level mapping function for Hss*. The variant of Hss referred to in Figure 1 uses the level mapping function, $\ell : [n] \to [R]$, where $R = 4\lceil \log(mM) \rceil$. Choose random $\ell_1, \ldots, \ell_R$, each $O(1)$-wise independent hash functions $[n] \to \{0, 1\}$. Define $\ell(i) = l$ iff $\ell_1(i) = 1, \ldots, \ell_l(i) = 1$ and either $l = R$ or $\ell_{l+1}(i) = 0$. This avoids the need for use of Nisan's PRG.

*Probability of* $\mathrm{NC}(H_g)$. Assume full independence of hash functions. For $i \in H_g$ and $l \in [s]$, let $w_{il} = 1$ if $i$ collides with some other item in $H_g$ in the table indexed $l$. Then, $q = \Pr[w_{il} = 1] =$

$1 - (1 - 1/(8C))^{|H_g|-1} \leq |H_g|/(8C) \leq 2/(100p^2)$. Let $W_i = \sum_{l=1}^{s}(1 - w_{il})$ be the number of tables where $i$ does not collide. Then, $\mathbb{E}[W_i] \geq s - 2s/(100p^2)$. By Chernoff's bounds, $W_i > s/2$ with probability at least $1 - e^{-s/12}$. By union bound, the probability that $W_i > s/2$ for each $i \in H_g$ is at least $1 - |H_g|e^{-s/12} = 1 - 2Bn^{-12} \geq 1 - 2n^{-11}$. Now, assuming $t$-wise independence, denote $q' = \Pr_t[w_{il} = 1]$. A standard inclusion-exclusion argument shows that $|q' - q| < 2\binom{n}{t}(q/n)^t < 2\left(\frac{e}{100p^2t}\right)^t$. This is at most $(1/16)q$ provided, $t \geq 3$. The above argument can be repeated using $q'$ instead of $q$. Combining, we have, $\text{NC}(H_g)$ holds with probability at least $1 - n^{-10}$.

**Proof of Corollary 2: Claim about bias.** Since, $\hat{f}_i \in |f_i| \pm \sigma$, $\hat{f}_i \in |f_i|(1 \pm 1/(9p))$. Further, $\frac{(x^p)^{(k)}}{k!} = \binom{p}{k}x^{p-k}$. This function is analytic in $\mathbb{R}$ if $p \geq k$ and otherwise is analytic in neighborhoods that do not contain 0. Since, $|f_i| \geq 1$ and $\hat{f}_i \in |f_i|(1 \pm 1/(9p))$, the function $x^p$ is analytic in the interval $[|f_i|, \hat{f}_i]$. Using Lemma 1, part (1), for some $\lambda' \in (|f_i|, \hat{f}_i)$, we have,

$$\left|\mathbb{E}[\vartheta_i] - |f_i|^p\right| = \binom{p}{k+1}\lambda'^{(p-k)}||f_i| - \hat{f}_i|^{k+1}$$

$$\leq |f_i|^p\left|\binom{p}{k+1}\right|\left(1 \pm 1/(9p)\right)^{p-k-1}(9p))^{-(k+1)} \tag{12}$$

where the plus sign is chosen if $p > k + 1$ and otherwise the minus sign is chosen.

*Case 1:* $k + 1 \geq p$. Then, $\binom{p}{k+1} = 0$ if $p$ is integral. Otherwise,

$$\left|\binom{p}{k+1}\right| = \frac{1}{(k+1)!} \cdot [p(p-1)\ldots(p-\lfloor p \rfloor)]\left[(\lfloor p \rfloor + 1 - p)(\lfloor p \rfloor + 2 - p)\cdots(k-p)\right]$$

$$= \frac{1}{(k+1)!} \cdot \left(\frac{\Gamma(p+1)}{|\Gamma(p - \lfloor p \rfloor)|}\right)\left(\frac{\Gamma(k+1-p)}{|\Gamma(\lfloor p \rfloor + 1 - p)|}\right)$$

$$\leq e(2/\pi)^{1/2}(k+1)^{-1/2} \cdot \frac{p^{p+1/2}(k-p)^{k-p+1/2}}{(k+1)^{k+1+1/2}}$$

$$= e(2/\pi)^{1/2}(k+1)^{-1/2}\left(\frac{p}{k+1}\right)^{p+1/2}\left(1 - \frac{p}{k+1}\right)^{k+1-p}$$

$$\leq e(2/\pi)^{1/2}(k+1)^{-1/2} .$$

Step 1 follows directly by rewriting $|p(p-1)\ldots(p-k)|$, Step 2 follows by writing each of the product expressions in terms of the Gamma function, Step 3 uses the Gamma reflection formula $|\Gamma(z)\Gamma(1-z)| = \pi/|\sin(\pi z)| \geq \pi$, Step 4 uses Stirling's approximation and Step 5 uses $p \leq k+1$. Applying to (12),

$$\left|\binom{p}{k+1}\right|\left(1 - 1/(9p)\right)^{p-k-1}(9p)^{-(k+1)} \leq e(2/\pi)^{1/2}(k+1)^{-1/2}e^{(k+1-p)/(9p)}(9p)^{-(k+1)}$$

$$\leq \frac{e^{8/9}\sqrt{2}(8p)^{-(k+1)}}{(\pi(k+1))^{1/2}} \leq (\epsilon/4)^{16} .$$

*Case 2:* $k+1 < p$. Then, $\left|\binom{p}{k+1}\right| \leq \frac{(pe)^{k+1}}{(k+1)^{k+1}}$. Applying to (12),

$$\left|\binom{p}{k+1}\right|\left(1+\frac{d}{c}\right)^{p-k-1}\left(\frac{d}{c}\right)^{k+1} \leq \frac{(pe)^{k+1}}{(9p(k+1))^{k+1}} e^{(p-k-1)/(9p)}$$
$$\leq e^{1/9}\left(\frac{e}{9(k+1)}\right)^{k+1} \leq 12^{-8}\epsilon^{12} .$$

The bias is bounded by $\max((\epsilon/4)^{16}, 12^{-8}\epsilon^{12}) \leq 12^{-8}\epsilon^{12}$. ∎

**Proof of Corollary 2: Claim about variance.** We have $\eta_i^2 = \sigma^2 + (\hat{f}_i - |f_i|)^2 \leq \sigma^2 + \sigma^2 = 2\sigma^2$. By Lemma 1, part (2),

$$\mathsf{Var}[\vartheta_i] \leq \Big(\sum_{v=1}^{k}\left|\binom{p}{v}\right|\hat{f}_i^{p-v}\eta_i^v\Big)^2 = \hat{f}_i^{2p-2}\eta_i^2\Big(\sum_{v=1}^{k}\left|\binom{p}{v}\right|\left(\frac{\eta_i}{\hat{f}_i}\right)^{v-1}\Big)^2 \quad (13)$$

Taking ratios of successive terms in the summation above, we have for $v \geq 1$, $p \geq 2$, $c \geq 9p$ and $c/d \geq 9p$

$$\frac{\left|\binom{p}{v+1}\right|\eta_i}{\left|\binom{p}{v}\right|\hat{f}_i} \leq \frac{|p-v|\sqrt{2}\sigma}{(v+1)(9p-1)\sigma} \leq \frac{p\sqrt{2}}{2(9p-1)} \leq 2^{-7/2} .$$

Substituting in (13),

$$\mathsf{Var}[\vartheta_i] \leq |f_i|^{2p-2}\left(1+\frac{1}{9p}\right)^{2p-2}\frac{(2\sigma^2)p^2}{(1-2^{-7/2})^2} \leq 3p^2\sigma^2|f_i|^{2p-2} . \quad ∎$$

*Averaged Taylor Polynomial Estimator.* We restate the construction of the estimator. Let $\{X_l\}_{l=1}^{s}$ be a family of independent and identical estimators with expectation $\mu$ and variance $\sigma^2$ and let $\lambda$ be an estimate of $\mu$. Let $s \geq 16k$ and $r = \Theta(s)$. Choose independently $r$ random permutations over $[s]$, denoted $\pi_1, \ldots, \pi_r$. For each permutation $\pi_j$, choose a random permutation $\pi'_j$ of the set $\pi_j([k]) = \{\pi_j(1), \ldots, \pi_j(k)\}$. Let $\tau_j$ denote $\pi'_j \circ \pi_j$. Order the elements of $\tau_j([k])$ in increasing order of the indices, that is, let $\tau_j([k]) = \{a_{j1}, a_{j2}, \ldots, a_{jk}\}$ where $a_{j1} < a_{j2} < \ldots < a_{jk}$. The averaged Taylor polynomial estimator $\bar{\vartheta}$ is defined as follows:

$$\vartheta_j = \sum_{v=0}^{k}\frac{\psi^{(v)}(\lambda)}{v!}\prod_{l=1}^{v}(X_{a_{jl}} - \lambda), \qquad \bar{\vartheta}(\psi,\lambda,k,r,s,\{X_l\}_{l=1}^{s}) = (1/r)\sum_{j=1}^{r}\vartheta_j .$$

The averaged polynomial estimator is also abbreviated as $\bar{\vartheta}(\psi, \lambda, k, r, s)$.

The following statement directly implies Lemma 3, by letting $r = 16s$.

**Lemma 13 (General form of Lemma 3.)** $\mathbb{E}\big[\bar{\vartheta}(\psi,\lambda,k,r)\big] = \mathbb{E}[\vartheta(\psi,\lambda,k)]$. *For* $\psi(x) = x^p$, $|\hat{f}_i - |f_i|| \leq \sigma$, $|f_i| \geq 9p\sigma$, $k \geq 144$, $s \geq 16k$,

$$\mathsf{Var}\big[\bar{\vartheta}(x^p,\hat{f}_i,k,r,s)\big] \leq p^2|f_i|^{2p-2}\sigma^2|f_i|^{2p-2}\sigma^2\big((3/r) + ((1+(72)^{-10})/s)\big) .$$

**Proof** The first statement follows by linearity of expectation.

$$\mathbb{E}\left[\bar{\vartheta}^2\right] = \frac{1}{r^2} \sum_{i=1}^{r} \mathbb{E}\left[\vartheta^2(S_i, \tau_i)\right] + \frac{2}{r^2} \sum_{1 \leq i < j \leq r} \mathbb{E}\left[\vartheta_i \vartheta_j\right] .$$

Let $Q_{ij}^{vv'}$ be the random variable that denotes the number of indices shared among the first $v$ and $v'$ positions of $\tau_i([k])$ and $\tau_j([k])$, that is, $Q_{ij}^{vv'} = \left|\{a_{i1}, \ldots, a_{iv}\} \cap \{a_{j1}, \ldots, a_{jv'}\}\right|$. Let $A_{iv}$ denote the set $\{a_{i1}, \ldots, a_{iv}\}$ and $A_{jv'}$ denote the set $\{a_{j1}, \ldots, a_{jv'}\}$. Fix an element $x \in A_{iv}$. Then, $x \in A_{jv'}$ provided, there exists $y \in [k]$ such that $\pi_j(y) = x$ and $\tau_j(y)$ is among the least $v'$ indices among $\tau_j([k])$. Hence,

$$\Pr\left[x \in A_{jv'}\right] = (k/s) \cdot (v'/k) = v'/s .$$

Sampling with replacement, we obtain an upper bound $\bar{Q}$ on $Q = Q_{ij}^{vv'}$. Then, $\bar{Q}$ is binomially distributed as $\text{Binom}(v; v'/s)$ and all moments of $\bar{Q}$ are at least as large as the corresponding moments of $Q$. Let $a_v(\lambda) = \psi^{(v)}(\lambda)/v!$. Then,

$$\mathbb{E}\left[\vartheta_i \vartheta_j\right] = \sum_{v,v'=0}^{k} a_v(\lambda) a_{v'}(\lambda) \mathbb{E}\left[\prod_{l=1}^{v}(X_{a_{il}} - \lambda) \prod_{l'=1}^{v'}(X_{a_{jl'}} - \lambda)\right]$$

$$= \sum_{v,v'=0}^{k} a_v(\lambda) a_{v'}(\lambda) \sum_{Q_{ij}^{vv'}=0}^{\min(v,v')} \prod_{t \in A_{iv} \cap A_{jv'}} \mathbb{E}\left[(X_t - \lambda)^2\right]$$

$$\times \prod_{t \in (A_{iv} \cup A_{jv'}) \setminus (A_{iv} \cap A_{jv'})} \mathbb{E}\left[X_t - \lambda\right] \Pr\left[Q_{ij}^{vv'} = q\right]$$

$$= \sum_{v,v'=0}^{k} a_v(\lambda) a_{v'}(\lambda) \sum_{q=0}^{\min(v,v')} \eta^{2q}(\mu - \lambda)^{v+v'-2q} \Pr\left[Q_{ij}^{vv'} = q\right]$$

$$\leq \sum_{v,v'=0}^{k} a_v(\lambda) a_{v'}(\lambda)(\mu - \lambda)^{v+v'} \mathbb{E}_{\bar{Q}_{ij}^{vv'}}\left[\frac{\eta^{2q}}{(\mu - \lambda)^{2q}}\right] \quad (14)$$

$$\leq \sum_{v,v'=0}^{k} a_v(\lambda) a_{v'}(\lambda)(\mu - \lambda)^{v+v'} \left(1 + v'/s\right)^v . \quad (15)$$

Step 1 follows from the definition of $\vartheta_i$ and $\vartheta_j$, Step 2 separates the shared variables $X_t$ with $t \in A_{iv} \cap A_{jv'}$ from the exclusive variables, Step 3 is a rewriting of the previous step, Step 4 replaces the distribution for $Q$ by the distribution $\bar{Q}$ which uses sampling with replacement. Since all moments of $\bar{Q}$ are at least as large as that of $Q$, this step follows. The final step uses a property of binomial distribution, namely, if $x$ has the distribution $\text{Binom}(n; p)$, then $\mathbb{E}\left[b^x\right] = (1 + (b-1)p)^n$. Since, $\bar{Q}_{ij}^{vv'}$ has the distribution $\text{Binom}(v; v'/s)$, we have

$$\mathbb{E}_{Q_{ij}^{vv'}}\left[\frac{\eta^{2q}}{(\mu - \lambda)^{2q}}\right] = \left[1 + \left(\frac{\eta^2}{(\mu - \lambda)^2} - 1\right)\frac{v'}{s}\right]^v = \left[1 + \frac{\sigma^2 v'}{(\mu - \lambda)^2 s}\right]^v .$$

We note that $\mathbb{E}[\vartheta_i \vartheta_j]$ is maximized when $\lambda$ is as far apart from $\mu$ as possible. Hence, (14) is maximized when $|\mu - \lambda|$ is as large as possible, which is $\sigma$. Thus, $\left[1 + \frac{\sigma^2 v'}{(\mu-\lambda)^2 s}\right]^v \leq \left(1 + v'/s\right)^v$, yielding (15).

We therefore have,

$$\mathbb{E}[\vartheta_i \vartheta_j] - \mathbb{E}[\vartheta_i]\mathbb{E}[\vartheta_j] = \sum_{v,v'=0}^{k} a_v(\lambda) a_{v'}(\lambda)(\mu - \lambda)^{v+v'} \left[(1 + v'/s)^v - 1\right]$$

$$= \sum_{v,v'=1}^{k} a_v(\lambda) a_{v'}(\lambda)(\mu - \lambda)^{v+v'} \left[(1 + v'/s)^v - 1\right] .$$

since, $(1 + v'/s)^v - 1$ is 0 if either $v$ or $v'$ is 0.

Divide the range of summation $1 \leq v, v' \leq k$ into regions $R_1 : 1 \leq v, v' \leq \lfloor\sqrt{k}\rfloor$ and $R_2 : \lfloor\sqrt{k}\rfloor + 1 \leq v, v' \leq k$. In the region $R_1$, $vv'/s \leq k/s \leq 1/16$.

$$\sum_{v,v'=1}^{\lfloor\sqrt{k}\rfloor} a_v(\lambda) a_{v'}(\lambda)(\mu - \lambda)^{v+v'} \left[(1 + v'/s)^v - 1\right]$$

$$= \sum_{v,v'=1}^{\lfloor\sqrt{k}\rfloor} a_v(\lambda) a_{v'}(\lambda)(\mu - \lambda)^{v+v'} \left[vv'/s + c(vv'/s)^2\right]$$

$$= \left[\sum_{v=1}^{\lfloor\sqrt{k}\rfloor} a_v(\lambda)(\mu - \lambda)^v (v/\sqrt{s} + cv^2/s)\right]^2, \quad \text{where, } |c| \leq 1/2. \quad (16)$$

Let $k_1 = \lfloor\sqrt{k}\rfloor$. Assuming that $\psi$ is analytic in $[\mu, \lambda]$,

$$\sum_{v=1}^{k_1} v a_v(\lambda)(\mu - \lambda)^v = (\mu - \lambda)\frac{d}{d\mu} \sum_{v=0}^{k_1} a_v(\lambda)(\mu - \lambda)^v = (\mu - \lambda)\frac{d}{d\mu}\left[a_0(\mu) - \sum_{t \geq k_1+1} a_t(\lambda)(\mu - \lambda)^t\right]$$

$$= a_1(\mu)(\mu - \lambda) - \sum_{t \geq k_1+1} t a_t(\lambda)(\mu - \lambda)^t \quad (17)$$

Similarly,

$$\sum_{v=1}^{k_1} v^2 a_v(\lambda)(\mu - \lambda)^v = \sum_{v=1}^{k_1} v a_v(\lambda)(\mu - \lambda)^v + (\mu - \lambda)^2 \frac{d^2}{d\mu^2} \sum_{v=0}^{k_1} a_v(\lambda)(\mu - \lambda)^v$$

$$= a_1(\mu)(\mu - \lambda) + a_2(\mu)(\mu - \lambda)^2 - \sum_{t \geq k_1+1} t^2 a_t(\lambda)(\mu - \lambda)^t \quad (18)$$

Let $\psi(x) = x^p$, $\mu = |f_i|$, $\lambda = \hat{f}_i$, $||f_i| - \hat{f}_i| \leq \sigma$, $|f_i| \geq 9p\sigma$ and $k \geq 144$. Let $b = \min(\lfloor p \rfloor, k_1 + 1)$. Then,

$$\sum_{t \geq k_1+1} t a_t(\lambda)(\mu - \lambda)^t = \sum_{t \geq k_1+1} t \binom{p}{t} \hat{f}_i^{p-t}(\mu - \lambda)^t$$

$$\leq \hat{f}_i^{p-k_1-1} |\mu - \lambda|^b \sum_{t \geq 0} (t + k_1 + 1) \left|\binom{p}{t + k_1 + 1}\right| (9p - 1)^{-t}$$

Taking ratios of successive terms of summation, we have for $t \geq 0$ and $k_1 = \sqrt{144} = 12$,

$$\frac{(t+k_1+2)\left|\binom{p}{t+k_1+2}\right|(9p-1)^{-t-1}}{(t+k_1+1)\left|\binom{p}{t+k_1+1}\right|(9p-1)^{-t}} = \frac{|p-(t+k_1+1)|}{(t+k_1+1)(9p-1)} \leq \frac{1}{17} \ .$$

Hence,

$$\sum_{t \geq k_1+1} t \binom{p}{t} \hat{f}_i^{p-t}(|f_i| - \hat{f}_i)^t \leq (k_1+1)\left|\binom{p}{k_1+1}\right| |\hat{f}_i^{p-k_1-1} \sigma^{k_1+1} \sum_{t \geq 0} 17^{-t}$$

$$\leq |f_i|^{p-1} e^{1/9} \sigma \left|\binom{p}{k_1+1}\right| (9p)^{-k_1} (18/17) \ .$$

For $k_1 + 1 \geq p$, $\left|\binom{p}{k_1+1}\right| \leq O((k+1)^{-1/2})$ as shown in the proof of Corollary 2. Otherwise, $\binom{p}{k_1+1} \leq (pe/(k_1+1))^{k_1+1}$. In either case,

$$(18/17)e^{1/9}(k_1+1)\left|\binom{p}{k_1+1}\right|(9p)^{-k_1} \leq (20/17)(18 \cdot 12/e)^{-12} \leq (72)^{-12} \ .$$

Similarly, it can be shown that

$$\sum_{t \geq k_1+1} t^2 \binom{p}{t} \hat{f}_i^{p-t}(|f_i| - \hat{f}_i)^t \leq (20/17)(k_1+1)^2(9p)^{-(k_1+1)}|f_i|^{p-1}\sigma \leq (72)^{-11}|f_i|^{p-1}\sigma$$

Rewriting (17) and (18) and using $k \geq 144$

$$\sum_{v=1}^{k_1} v \binom{p}{v} \hat{f}_i^{p-v}(|f_i| - \hat{f}_i)^v \leq p|f_i|^{p-1}\sigma(1 + (72)^{-12})$$

$$\sum_{v=1}^{k_1} v^2 \binom{p}{v} \hat{f}_i^{p-v}(|f_i| - \hat{f}_i)^v \leq p|f_i|^{p-1}\sigma(1 + (72)^{-11}) \ .$$

Combining, and substituting in (16) we have

$$\text{RHS of (16)} \leq (1 + 2(72)^{-11})p|f_i|^{p-1}\sigma)^2/s \ . \tag{19}$$

We now consider the second region $R_2 : \lfloor \sqrt{k} \rfloor + 1 \leq v, v' \leq k$. Then,

$$\sum_{v,v'=k_1+1}^{k} a_v(\lambda) a_{v'}(\lambda)(\mu - \lambda)^{v+v'}\left[(1 + v'/s)^v - 1\right]$$

$$= \sum_{v,v'=k_1+1}^{k} \binom{p}{v}\binom{p}{v'} \hat{f}_i^{2p-(v+v')}(\alpha\sigma)^{v+v'}(1 + v'/s)^v$$

$$- \left[\sum_{v=k_1+1}^{k} \binom{p}{v} \hat{f}_i^{p-v}(\alpha\sigma)^v\right]^2, \quad \text{where, } |\alpha| < 1.$$

$$\leq \sum_{v,v'=k_1+1}^{k} \binom{p}{v}\binom{p}{v'} \hat{f}_i^{2p-(v+v')}(\alpha\sigma)^{v+v'}(1 + v'/s)^v$$

$$= \sum_{v,v'=k_1+1}^{k} \binom{p}{v}\binom{p}{v'} \hat{f}_i^{2p-(v+v')}(\alpha\sigma)^{v+v'} e^{\alpha'_{v'}(v+v')k/(2s)} \tag{20}$$

since for each value of $v'$, there is a value $\alpha'_{v'} \in [0,1]$ such that

$$(1+v'/s)^v = e^{\alpha'_{v'}(v+v')k/(2s)} \quad \text{since, } v, v' \leq k.$$

From (20), it follows that there exists a value of $\beta \in [0,1]$ such that

$$\sum_{v,v'=k_1+1}^{k} \binom{p}{v}\binom{p}{v'} \hat{f}_i^{2p-(v+v')}(\alpha\sigma)^{v+v'} e^{\alpha'_{v'}(v+v')k/(2s)}$$
$$= \sum_{v,v'=k_1+1}^{k} \binom{p}{v}\binom{p}{v'} \hat{f}_i^{2p-(v+v')}(\alpha\sigma)^{v+v'} e^{\beta(v+v')k/(2s)} \ .$$

Therefore, for some $0 \leq |\alpha|, \beta \leq 1$ and $f_{i1}, f_{i2} \in \hat{f}_i \pm \alpha\sigma e^{\beta k/(2s)}$, we have

$$\sum_{v,v'=k_1+1}^{k} a_v(\lambda)a_{v'}(\lambda)(\mu-\lambda)^{v+v'}\left[(1+v'/s)^v - 1\right]$$
$$= \sum_{v,v'=k_1+1}^{k} \binom{p}{v}\binom{p}{v'} \hat{f}_i^{2p-(v+v')}(\alpha\sigma)^{v+v'} e^{\beta(v+v')k/(2s)}$$
$$= \Big[\sum_{v=k_1+1}^{k} \binom{p}{v} \hat{f}_i^{p-v}(\alpha\sigma e^{\beta k/(2s)})^v\Big]^2$$
$$\leq |f_i|^{2p-2}\sigma^2(72)^{-24} \ . \tag{21}$$

Combining (19) and (21), and since $k \geq 144$, we have,

$$\mathbb{E}[\vartheta_i\vartheta_j] - \mathbb{E}[\vartheta_i]\mathbb{E}[\vartheta_j] \leq (1+2(72)^{-11})^2 p^2|f_i|^{2p-2}\sigma^2/s + |f_i|^{2p-2}\sigma^2(72)^{-24}$$
$$\leq (1+(72)^{-10})p^2|f_i|^{2p-2}\sigma^2/s \ .$$

Therefore, for any fixed $i, j$ with $i \neq j$, we have

$$\mathsf{Var}[\bar{\vartheta}] = \frac{1}{r}\mathsf{Var}[\vartheta_i] + \frac{2}{r^2}\binom{r}{2}\left(\mathbb{E}[\vartheta_i\vartheta_j] - \mathbb{E}[\vartheta_i]\mathbb{E}[\vartheta_j]\right)$$
$$\leq |f_i|^{2p-2}\sigma^2\left((3p^2/r) + ((1+(72)^{-10})p^2/s\right) \ . \quad \blacksquare$$